\documentclass[12pt]{article}
\usepackage{amsmath,amssymb,amscd,amsbsy,amsgen,amsopn,amstext,
amsxtra}

\usepackage[dvips]{graphicx}

\begin{document}

\begin{flushright}
\end{flushright}

\vskip 0.5 truecm

\begin{center}
{\Large{\bf Aspects of universally valid Heisenberg uncertainty relation}}
\end{center}
\vskip .5 truecm
\centerline{\bf  Kazuo Fujikawa$^1$ and Koichiro Umetsu$^{2}$}
\vskip .4 truecm
\centerline {\it $^1$ Mathematical Physics Laboratory, RIKEN 
Nishina Center,}
\centerline {\it Wako 351-0198, 
Japan}
\vspace{0.3cm}
\centerline {\it $^2$ Maskawa Institute for Science and Culture, }
\centerline {\it Kyoto Sangyo University, Kyoto 603-8555, Japan }
\vskip 0.5 truecm

\makeatletter
\makeatother

\begin{abstract}
A numerical illustration of a universally valid Heisenberg uncertainty relation, which was proposed recently, is presented by using the experimental data on spin-measurements by J. Erhart, et al.[ Nature Phys. {\bf 8}, 185 (2012)]. This uncertainty relation is closely related to a modified form of the Arthurs-Kelly uncertainty relation which is also tested by the spin-measurements. The universally valid Heisenberg uncertainty  relation always holds, but both  the modified Arthurs-Kelly uncertainty  relation and   Heisenberg's  error-disturbance relation proposed by Ozawa, which was analyzed in the original experiment, fail in the present context of spin-measurements, and the cause of their failure is identified with the assumptions of unbiased measurement and disturbance. It is also shown that all the universally valid uncertainty relations are derived from Robertson's relation and thus the essence of the uncertainty relation is exhausted by Robertson's relation as is widely accepted.
\end{abstract}

 
\section{Introduction}   
The uncertainty relation forms the basis of the entire quantum theory~\cite{heisenberg, kennard, robertson}. The original formulation by Heisenberg~\cite{heisenberg} is based on a thought experiment and emphasizes  measurement processes.
On the other hand, the formulations by Kennard~\cite{kennard} and Robertson~\cite{robertson}, which are mathematically well-defined, usually evaluate only the intrinsic quantum fluctuations without explicit  reference to measurement processes. The formulations by Kennard and Robertson are widely accepted as the mathematical expression of the Heisenberg uncertainty relation. In the spirit of the original formulation of Heisenberg, however, it may be desirable to mention explicitly the effects of measurement in the formulation of the uncertainty relation. Our recent interest in the uncertainty relations arises from 
the beautiful experiment~\cite{hasegawa} which invalidated a general form of
Heisenberg's error-disturbance relation proposed in the past~\cite{ozawa1}.

The measurement process was directly incorporated into the uncertainty relation by Arthurs and Kelly~\cite{arthurs} and Arthurs and Goodman~\cite{arthurs2}. Their relation is based on 
the assumption of joint unbiased measurements, which has no general proof.  A different way to incorporate the measurement process 
into the uncertainty relation was found by Ozawa who derived a universally valid error-disturbance relation by combining a triangle inequality with Robertson's relations~\cite{ozawa1, ozawa2} and proposed  a Heisenberg's error-disturbance relation which was recently invalidated~\cite{hasegawa}. In the present paper, those relations based on the positive definite Hilbert space and natural commutator algebra are referred to as "universally valid" since they are expected to be valid as long as quantum mechanics is valid. Recently one of the present authors proposed a universally valid Heisenberg uncertainty relation~\cite{fujikawa}, which was intended to provide support for the original idea of Heisenberg, as an alternative to an invalid  Heisenberg's error-disturbance relation proposed by Ozawa~\cite{ozawa1}, and the proposed relation revealed an unexpected and interesting generalization of the Arthurs-Kelly relation. This proposed universally valid Heisenberg relation, which is mathematically a further combination of Robertson's relation and  Ozawa's universally valid error-disturbance relation, is free of the assumptions of unbiased measurement and disturbance and its validity is at the same level as the standard Kennard and Robertson relations~\cite{kennard, robertson}. The message in~\cite{fujikawa} is that the original idea of Heisenberg is correct despite the negative statements in~\cite{hasegawa, ozawa1}.  

The purpose of the present paper is to show that the universally valid Heisenberg uncertainty relation is a natural candidate of the manifestation of Heisenberg's original idea by clarifying its characteristic features  with future practical applications in mind.  
 We first numerically illustrate, on the basis of the experimental data~\cite{hasegawa}, the satisfactory features of the universally valid Heisenberg uncertainty relation  but the failure of a closely related modified form of Arthurs-Kelly uncertainty relation in the present context. 
We then show that the Heisenberg's error-disturbance relation of Ozawa is derived on the same basis as the modified Arthurs-Kelly relation by assuming the unbiased measurement and disturbance, and the origin of their failure is discussed. It is also shown that all the uncertainty relations discussed in this paper including the universally valid Heisenberg relation are derived from Robertson's relation which is widely accepted as the Heisenberg uncertainty relation. We  briefly comment on the implications of the notion of precise measurement on the formulation of uncertainty relations. 

\section{Universally valid Heisenberg uncertainty relation}

A universally valid Heisenberg uncertainty relation, which supports the original idea of Heisenberg, has been recently proposed~\cite{fujikawa}. This relation assumes the form
\begin{eqnarray}
\bar{\epsilon}(A)\bar{\eta}(B)\geq |\langle [A,B]\rangle|
\end{eqnarray}
where
\begin{eqnarray}
\bar{\epsilon}(A)&\equiv&\epsilon(A)+\sigma(A)\nonumber\\
&=&\langle (M^{out}-A)^{2}\rangle^{1/2}+\langle (A-\langle A\rangle)^{2}\rangle^{1/2},\nonumber\\
\bar{\eta}(B)&\equiv&\eta(B)+\sigma(B)\nonumber\\
&=&\langle (B^{out}-B)^{2}\rangle^{1/2}+\langle (B-\langle B\rangle)^{2}\rangle^{1/2}.
\end{eqnarray}
The motivation for writing the relation (1), as an alternative  to the proposal of a different invalid relation as Heisenberg's error-disturbance relation by Ozawa~\cite{ozawa1}, and the close similarity of (1) with another basic relation, namely the Arthurs-Kelly relation, are explained in detail in~\cite{fujikawa}.  
We here work in the Heisenberg representation and those variables without any suffix stand for the initial variables. $M^{out}$ stands for the meter variable after the measurement of $A$, and 
$B^{out}$ stands for the conjugate variable after the measurement of $A$. It was suggested in Ref.~\cite{fujikawa} that  $\bar{\epsilon}(A)$ is called  the "inaccuracy" in the measured values of the variable $A$ (to avoid the confusion with the commonly used "error" for $\epsilon(A)$), and  
$\bar{\eta}(B)$ is called the inevitable "fluctuation" in the conjugate variable $B$ (to avoid the confusion with "disturbance" for $\eta(B)$) after the measurement of $A$. The quantity 
\begin{eqnarray}
\epsilon(A)=\langle (M^{out}-A)^{2}\rangle^{1/2}
\end{eqnarray}
is commonly referred to as "error" in the measurement of $A$, and 
\begin{eqnarray}
\eta(B)=\langle (B^{out}-B)^{2}\rangle^{1/2}
\end{eqnarray}
as "disturbance" in the variable $B$ after the measurement of $A$. 
\begin{eqnarray}
\sigma(A)=\langle (A-\langle A\rangle)^{2}\rangle^{1/2}
\end{eqnarray}
is the standard deviation.

The relation (1) has a clear physical meaning only for the repeated measurements of similarly
prepared samples in quantum mechanics, as is emphasized by the explicit presence of standard deviations in the relation~\footnote{All the universally valid relations are derived from Robertson's relation, as is explained later in Section 3, and thus the quantum average is always included. However, the explicit presence of the standard deviations which represent intrinsic quantum fluctuations, in addition to the measurement-related "error" or "disturbance", emphasizes the sample average.}. For example, the "precise" measurement 
defined by the vanishing error $\epsilon(A)=\langle (M^{out}-A)^{2}\rangle^{1/2}=0$~\cite{ozawa1} suggests
\begin{eqnarray}
M^{out}|\psi\rangle\otimes|\xi\rangle=A|\psi\rangle\otimes|\xi\rangle
\end{eqnarray}
where $|\psi\rangle$ stands for the state vector of the system and $|\xi\rangle$ is a suitable state vector of the measuring apparatus~\cite{ozawa2, neumann} in the Heisenberg representation.
We thus have the relation between standard deviations
\begin{eqnarray}
\sigma(M^{out})=\sigma(A)
\end{eqnarray}
defined for the state $|\psi\rangle\otimes|\xi\rangle$,
namely, even for the "precise" measurement we have the standard deviation $\sigma(M^{out})=\sigma(A)$ of the measurement apparatus if one performs the {\em repeated} measurements in quantum mechanics. This property may
justify the identification of $\bar{\epsilon}(A)$ in (2) as the "inaccuracy" in the measured values  of $A$ even for the  "precise" measurement. Similarly, one can identify $\bar{\eta}(B)$ in (2) as the inevitable "fluctuation" of the conjugate variable $B$ after the measurement of $A$, if one repeats the measurements of similarly prepared samples; even if $\eta(B)=\langle (B^{out}-B)^{2}\rangle^{1/2}=0$, namely, for $B^{out}|\psi\rangle\otimes|\xi\rangle=B|\psi\rangle\otimes|\xi\rangle
$, we have the fluctuation $\sigma(B^{out})=\sigma(B)$ in the variable $B^{out}$.

Obviously, the inaccuracy $\bar{\epsilon}(A)$ has no clear meaning for a single measurement for a broad $\sigma(A)$, for example, and only after the repeated measurements the inaccuracy 
$\bar{\epsilon}(A)=\epsilon(A)+\sigma(A)$ gives a criterion of the "good" measurement of $A$.  
The physical meaning of the universally valid Heisenberg uncertainty relation (1) is that a good
measurement of $A$ requires a well localized state with small $\sigma(A)$ and the precise measurement with small error $\epsilon(A)$. This good measurement cannot be achieved for the small fluctuation $\bar{\eta}(B)=\eta(B)+\sigma(B)$ in the conjugate variable $B$, namely, for the well localized 
conjugate variable with small $\sigma(B)$ and negligible disturbance in the conjugate variable with 
small $\eta(B)$. The Heisenberg uncertainty relation in the form of the universally valid Heisenberg relation is thus a negative statement, and the relation 
(1) represents the inevitable limits dictated by quantum mechanics. The universally valid Heisenberg relation is constructed~\cite{fujikawa} by a combination of a triangle inequality  with Robertson's relations~\cite{robertson}
\begin{eqnarray}
\sigma(A)\sigma(B)\geq \frac{1}{2}|\langle [A,B]\rangle|
\end{eqnarray}
for general hermitian operators $A$ and $B$. Robertson's relations are  based on standard deviations and thus may appear to have no direct connection with  measurement processes, but the notions of "error" and "disturbance" are incorporated by the replacements of  standard deviations for suitable operators by error or disturbance in a formal manner~\cite{ozawa1}. It is hoped that our intuitive understanding of the physical meanings of inaccuracy $\bar{\epsilon}(A)$ and fluctuation $\bar{\eta}(B)$, which are introduced in the universally valid Heisenberg uncertainty relation in a formal manner, will be deepened in the course of its practical applications.

The relation (1) is similar to the Arthurs-Kelly relation~\cite{arthurs,arthurs2},
\begin{eqnarray}
\sigma(M^{out})\sigma(B^{out})&=&\{\langle (M^{out}-A)^{2}\rangle+\langle (A-\langle A\rangle)^{2}\rangle \}^{1/2}\nonumber\\
&&\times\{\langle (B^{out}-B)^{2}\rangle+\langle (B-\langle B\rangle)^{2}\rangle \}^{1/2}\nonumber\\
&\geq& |\langle [A,B]\rangle|
\end{eqnarray}
which also emphasizes the repeated measurements of similarly prepared samples in quantum mechanics by the explicit presence of standard deviations $\sigma(A)$ and $\sigma(B)$. To be precise, the relation in (9) is based on the unbiased measurement and disturbance while the original Arthurs-Kelly relation is based on the joint unbiased measurements~\cite{arthurs,arthurs2}. There is however a logical relationship between these two relations~\cite{ozawa1} if one assumes that the measurement of $A$ using the apparatus ${\cal A}$ is
followed immediately by a measurement of the observable $B$ using a noiseless measuring apparatus ${\cal B}$. Then, combining the two apparatus, we may arrive at the relation (9). We find this argument convincing, but the notion of unbiased disturbance is subtle and for this reason, we refer to the relation (9) as a modified form of Arthurs-Kelly
relation in the following. This modified Arthurs-Kelly relation is important to understand the physical meaning of our universally valid Heisenberg uncertainty relation (1), since both relations have the common lower bound and both of them contain a combination of the error and standard deviation or a combination of the disturbance and standard deviation,  although the different combinations in (1) and (9); $\bar{\epsilon}(A)$ and $\sigma(M^{out})$, for example,  agree for very small $\epsilon(A)$ or very small $\sigma(A)$. The quantities  $\sigma(M^{out})$ and $\sigma(B^{out})$ have very clear physical meanings and they will help deepen our understanding of the physical meanings of "inaccuracy" $\bar{\epsilon}(A)$ and "fluctuation" $\bar{\eta}(B)$.

We note the relations
\begin{eqnarray}
\sigma(M^{out})&=&\{\langle (M^{out}-A)^{2}\rangle+\langle (A-\langle A\rangle)^{2}\rangle \}^{1/2}\nonumber\\
&\leq&\{\langle (M^{out}-A)^{2}\rangle^{1/2}+\langle (A-\langle A\rangle)^{2}\rangle^{1/2}\},
\end{eqnarray}
and thus $\bar{\epsilon}(A)\geq \sigma(M^{out})$, and 
\begin{eqnarray}
\sigma(B^{out})&=&\{\langle (B^{out}-B)^{2}\rangle+\langle (B-\langle B\rangle)^{2}\rangle \}^{1/2}\nonumber\\
&\leq&\{\langle (B^{out}-B)^{2}\rangle^{1/2}+\langle (B-\langle B \rangle)^{2}\rangle^{1/2}\}
\end{eqnarray}
and thus $\bar{\eta}(B)\geq \sigma(B^{out})$. We thus have  
\begin{eqnarray}
\bar{\epsilon}(A)\bar{\eta}(B)\geq \sigma(M^{out})\sigma(B^{out}), 
\end{eqnarray}
namely, the universally valid Heisenberg uncertainty relation (1), which is defined without referring to the unbiased measurement and disturbance,  holds more generally than the modified Arthurs-Kelly relation (9) which is based on the assumptions of  unbiased measurement and disturbance~\cite{appleby}. 
One may even regard the universally valid Heisenberg relation as a universally valid version of the modified Arthurs-Kelly relation.

We later show that Heisenberg's error-disturbance relation proposed by Ozawa~\footnote{This relation, which was proposed to be identified with the original Heisenberg's error-disturbance relation by Ozawa~\cite{ozawa1}, is called "a general form of Heisenberg's error-disturbance relation " in~\cite{hasegawa}. There appears to be no consensus on the naming of this relation. We tentatively call it "Heisenberg's error-disturbance relation proposed by Ozawa"  or "Heisenberg's error-disturbance relation of Ozawa", but we have no intention to advocate a specific naming.},
\begin{eqnarray}
\epsilon(A)\eta(B)\geq \frac{1}{2}|\langle[A,B]\rangle|,
\end{eqnarray}
which was invalidated by the recent experiment~\cite{hasegawa},
is based on the  precisely same set of assumptions, namely, unbiased measurement and disturbance.

See Ref.~\cite{fujikawa} for the derivation of (9) on the basis of the unbiased measurement and disturbance and Robertson's relation; the relation (13) is also used in this derivation. In contrast, the derivation of (1) is free of the assumptions of unbiased measurement and disturbance.

In Fig.1, we compare the relation (1) with the relation (9) and with the recent experimental data
in~\cite{hasegawa}. We use the theoretical values of $\epsilon(A)=||(E_{\phi}(+1)-E_{\phi}(-1)-\sigma_{x})|\psi\rangle||=2\sin(\frac{\phi}{2})$, $\eta(B)=\sqrt{2}||[E_{\phi}(+1),\sigma_{y}]|\psi\rangle||=\sqrt{2}\cos\phi$, $\sigma(A)=1$ and 
$\sigma(B)=1$ with $\phi$ standing for the detuning angle  $\sigma_{\phi}=\cos\phi\sigma_{x}+\sin\phi\sigma_{y}$ defined in~\cite{hasegawa}, and projection operators $E_{\phi}(\pm 1)=(1\pm \sigma_{\phi})/2$. Here we choose $A=\sigma_{x}$ and $B=\sigma_{y}$ and $|\psi\rangle=|+z\rangle$. (For simplicity, $\hbar/2$ is omitted for observables of each spin component~\cite{hasegawa}.) As for the experimental data, which is compared with the theoretical predictions, we read off the values given in Figures 4 and 5 in ~\cite{hasegawa}.  

\begin{figure}[h]
  \begin{center}
    \includegraphics[width=100mm]{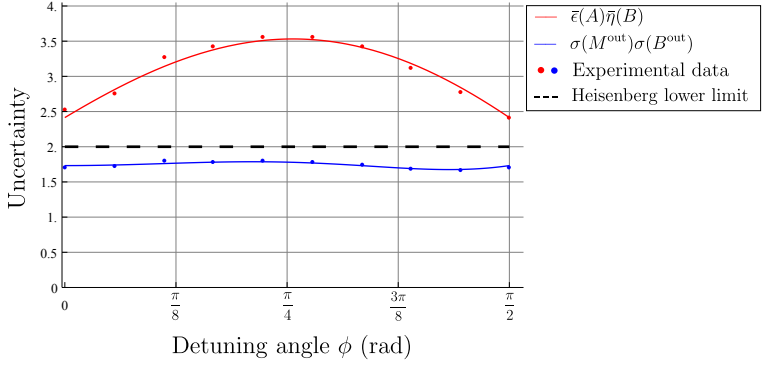}
  \end{center}
  \caption{{\bf Illustration of the universally valid Heisenberg uncertainty relation in (1) (red line) and the modified Arthurs-Kelly relation in (9) (blue line):}~The theoretical predictions in (14) and (15) are compared to the experimental data given in Figures 4 and 5 of~\cite{hasegawa}. 
For simplicity, $\hbar/2$ is omitted for observables of each spin component.}
  \label{.eps}
\end{figure}

We see in Fig.1 that our universally valid Heisenberg uncertainty relation in (1) written in the form,
\begin{eqnarray}
(2\sin(\frac{\phi}{2})+1)(\sqrt{2}\cos\phi+1)\geq 2, 
\end{eqnarray}
is always satisfied by the experimental data, while the modified Arthurs-Kelly relation in (9) written in the form, 
\begin{eqnarray}
\{(4\sin^{2}(\frac{\phi}{2})+1)(2\cos^{2}\phi+1)\}^{1/2}\geq 2,
\end{eqnarray}
is violated for all the detuning angles both theoretically and experimentally:
The good agreement of theoretical and experimental values in Figures 4 and 5 of Ref.~\cite{hasegawa} shows that the experimental values of $\epsilon(A)$, $\eta(B)$, $\sigma(A)$ and $\sigma(B)$ agree well with theoretical estimates  $\epsilon(A)=2\sin(\frac{\phi}{2})$, $\eta(B)=\sqrt{2}\cos\phi$, $\sigma(A)=1$ and $\sigma(B)=1$ in~\cite{hasegawa, ozawa2}. Naturally, our Fig.1 also shows the agreement of experimental values with the analytical formulas in (14) and (15), and the comparison with the lower bound $2$ tests the uncertainty relations. 

The spin measurement~\cite{hasegawa} thus illustrates for the first    time a clear failure of the modified Arthurs-Kelly relation and thus, with the qualifications already stated, suggests the possible complications with the Arthurs-Kelly relation itself which has been otherwise considered to be valid~\cite{she, yuen, yamamoto}.  This failure of the modified Arthurs-Kelly relation is traced to the failure of the unbiased measurement and disturbance and thus it is related to the failure of the relation (13), but these two relations (9) and (13) are not physically identical. The failure of the relation (13) casts doubt on the derivation of the modified Arthurs-Kelly relation in (9)  but it does not necessarily imply the (complete) failure of the modified Arthurs-Kelly relation.

In comparison, a universally valid error-disturbance relation 
\begin{eqnarray}
\epsilon(A)\eta(B)+\sigma(A)\eta(B)+\epsilon(A)\sigma(B)\geq \frac{1}{2}|\langle [A,B]\rangle|
\end{eqnarray}
was proposed by Ozawa~\cite{ozawa1,ozawa2} by combining a triangle inequality with Robertson's relations (8). This relation played a pivotal role in the proposal of the two Heisenberg-type relations, namely, to derive the relation (13) by throwing away two terms and to derive our relation (1) by adding a term.
Physically, our relation is close to the modified Arthurs-Kelly relation (9) with emphasis on repeated measurements by the explicit presence of  standard deviations in a factored form for two conjugate variables, as was already discussed. We emphasize that both Ozawa's error-disturbance relation (16) and our relation (1) are constructed by the combinations of Robertson's relations (8) with the suitable replacements of standard deviations $\sigma$ by the error
$\epsilon$ or disturbance $\eta$, and thus the essence of all the universally valid uncertainty relations is exhausted by Robertson's relation. This fact is explained in Section 3 later.

\begin{figure}[h]
  \begin{center}
    \includegraphics[width=120mm]{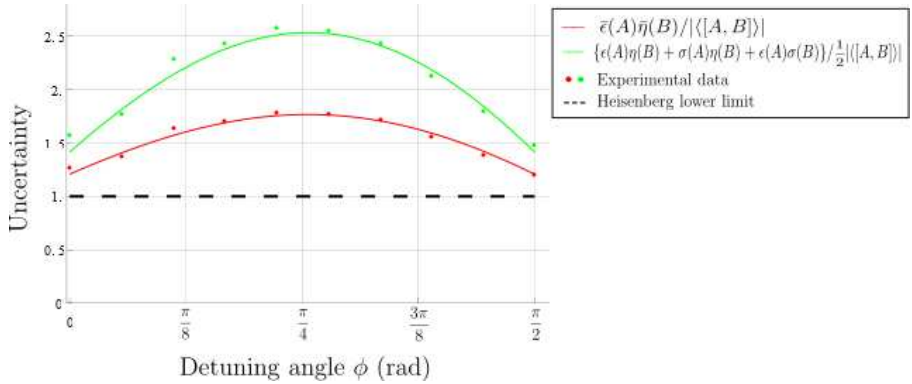}
  \end{center}
  \caption{{\bf Comparison of the universally valid Heisenberg uncertainty relation in (1) (red line) with Ozawa's universally valid error-disturbance uncertainty relation in (16) (green line):}~Both relations with experimental data~\cite{hasegawa} are consistent with the uncertainty bound, and the relation in (1) comes closer to the lower bound  if one normalizes the lower bound to be unity, of which meaning is explained in the body of the paper. The factor $\hbar/2$ is omitted for observables of each spin component.}
  \label{.eps}
\end{figure}

In Fig.2, we compare our relation (1) with Ozawa's relation (16); both are shown to be consistent with the experimental data by following the analysis in~\cite{hasegawa}. There is no absolute way to compare the two relations (1) and (16), but in our comparison we normalized the uncertainty relations by  their lower bounds , namely, we show 
\begin{eqnarray}
\bar{\epsilon}(A)\bar{\eta}(B)/|\langle [A,B]\rangle|
\end{eqnarray}
instead of $\bar{\epsilon}(A)\bar{\eta}(B)$, for example. The lower bounds are thus automatically normalized to unity. 
Mathematically, this procedure corresponds to the comparison of two numbers defined by the difference between left- and right-hand sides of each inequality normalized by the right-hand side, namely, 
\begin{eqnarray}
\{\bar{\epsilon}(A)\bar{\eta}(B)- |\langle [A,B]\rangle|\}/|\langle [A,B]\rangle|
\end{eqnarray}
and 
\begin{eqnarray}
\{\epsilon(A)\eta(B)+\sigma(A)\eta(B)+\epsilon(A)\sigma(B)- \frac{1}{2}|\langle [A,B]\rangle|\}/\frac{1}{2}|\langle [A,B]\rangle|,
\end{eqnarray}
respectively. We compare the ratios of the deviations from the lower bounds to the lower bounds themselves; for example, Fig.2 shows that our relation deviates from its lower bound by about 75 percent and Ozawa's relation deviates from its lower bound by about 150 percent for the detuning angle $\phi=\pi/4$, respectively.
Physically, this procedure may provide one of the sensible ways to compare the two 
different inequalities when one fits the left- and right-hand sides of the inequalities with measured data.
With those qualifications in mind, it is still interesting to see in Fig.2 that our relation comes closer to the lower bound in the present example.
 
In this connection, we note that Ozawa's universally valid error-disturbance relation (16) is naturally compared with the Heisenberg's error-disturbance relation (13) proposed by him just as the universally valid Heisenberg uncertainty relation (1) is naturally compared with the modified Arthurs-Kelly relation (9), although the universally valid Heisenberg relation was originally presented as an
alternative to the Heisenberg's error-disturbance relation of Ozawa.

\section{All the uncertainty relations from Robertson's relation}

In this section, we show that all the uncertainty relations we use in this paper are derived from Robertson's relation and, in particular, the derivation of the relation (13) on the basis of the extra assumptions of unbiased measurement and disturbance. We then discuss the possible origin of the failure of the relations (9) and (13). We also briefly comment on the implications of the notion of precise measurement on the Heisenberg-type uncertainty relations.

We start with the universally valid Robertson's relation~\cite{robertson}   
\begin{eqnarray}
\sigma(M^{out}-A)\sigma(B^{out}-B)\geq \frac{1}{2}|\langle [M^{out}-A,B^{out}-B]\rangle|
\end{eqnarray}
which may be formally re-written in the present notations of $\epsilon(A)$ and $\eta(B)$ by noting $\epsilon(A)\geq \sigma(M^{out}-A)$ and $\eta(B)\geq \sigma(B^{out}-B)$ as~\cite{ozawa1}
\begin{eqnarray}
\epsilon(A)\eta(B)\geq \frac{1}{2}|\langle [M^{out}-A,B^{out}-B]\rangle|
\end{eqnarray}
which is still universally valid~\footnote{This formal replacement of the standard deviations by $\epsilon$ or $\eta$ is common to (1) and (16). The quantity 
$\sigma(B^{out}-B)$ in Robertson's relation (20), for example, is originally defined as an average
of the operator $B^{out}-B$ by using the state $|\psi\otimes\xi\rangle$ for any given $B^{out}$, in principle independently of the measurement of $A$. But after the above replacement, the quantity $\eta(B)$ is interpreted as a disturbance caused by the measurement of $A$ by assigning a specific time development to $B^{out}$.}. 

We next note that by assuming the unbiased measurement and disturbance we have
\begin{eqnarray}
|\langle [M^{out}-A,B^{out}-B]\rangle|&=&
|\langle [M^{out}-A,B^{out}]\rangle|\nonumber\\
&=&|\langle [A,B^{out}]\rangle|\nonumber\\
&=&|\langle [A,B^{out}-B+B]\rangle|\nonumber\\
&=&|\langle [A,B]\rangle|
\end{eqnarray}
where we used the properties of the unbiased measurement and disturbance,
$\langle [M^{out}-A,B]\rangle=\langle [A,B^{out}-B]\rangle=0$. To understand these relations, one may note the following identity shown in Appendix of~\cite{appleby}
\begin{eqnarray}
\langle\psi\otimes\xi|{\cal A}|\psi^{\prime}\otimes\xi\rangle&=&
\frac{1}{4}\{
\langle(\psi+\psi^{\prime})\otimes\xi|{\cal A}|(\psi+\psi^{\prime})\otimes\xi\rangle\nonumber\\
&&-\langle(\psi-\psi^{\prime})\otimes\xi|{\cal A}|(\psi-\psi^{\prime})\otimes\xi\rangle\nonumber\\
&&-i\langle(\psi+i\psi^{\prime})\otimes\xi|{\cal A}|(\psi+i\psi^{\prime})\otimes\xi\rangle\nonumber\\
&&+i\langle(\psi-i\psi^{\prime})\otimes\xi|{\cal A}|(\psi-i\psi^{\prime})\otimes\xi\rangle\}
\end{eqnarray}
with ${\cal A}=M^{out}-A$ and $\psi^{\prime}=B\psi$, for example. One may then recall the definition of the unbiased measurement $\langle\psi\otimes\xi|M^{out}-A|\psi\otimes\xi\rangle=0$ for {\em all} $\psi$~\cite{appleby}, and similarly the unbiased disturbance.
We also used the commutation relation $\langle [M^{out},B^{out}]\rangle=0$ of canonically non-conjugate variables.
We thus obtain the Heisenberg's  error-disturbance  relation of Ozawa (13) from the universally valid Robertson's relation (20) by assuming the unbiased measurement and disturbance.  See also Ref.~\cite{ozawa3} for an alternative derivation of (13).

The difficulty of the relation (13) is understood by 
noting 
\begin{eqnarray}
\langle (M^{out}-A)^{2}\rangle\langle (B^{out}-B)^{2}\rangle = 0
\end{eqnarray}
for the bounded operator $B$ such as the spin variable  for the "precise" measurement $\langle (M^{out}-A)^{2}\rangle=0$ regardless of the value of $\frac{1}{2}|\langle [A,B]\rangle|$~\cite{ozawa2}. See also the recent experimental result~\cite{hasegawa}.

On the other hand, the universally valid relation (21) is valid even for the precise measurement which implies $M^{out}|\psi\rangle\otimes|\xi\rangle=A|\psi\rangle\otimes|\xi\rangle$ in (6), and thus the right-hand side of (21) also vanishes for the precise measurement. (In this analysis, the finite $\eta(B)$ or $||B^{out}-B||<\infty$ is important.) Obviously, the failure of the Heisenberg's error-disturbance relation of Ozawa (13)
arises from the invalid assumptions of unbiased measurement and disturbance for bounded operators such as spin variables. In connection with future experimental studies, it is an interesting question if the assumptions of  unbiased measurement and disturbance are valid for unbounded operators.
\\

It is now shown that all the uncertainty relations are derived from  Robertson's relation (20). For this purpose, we rewrite  Robertson's relation (20) using the triangle inequality (in the two-dimensional space of complex numbers) in the form
\begin{eqnarray}
&&\sigma(M^{out}-A)\sigma(B^{out}-B)\nonumber\\
&&\geq \frac{1}{2}|\langle [-A,B^{out}-B]+[M^{out}-A,-B]-[-A,-B]\rangle|\nonumber\\
&&\geq \frac{1}{2}\{|\langle [A,B]\rangle|-|\langle [A,B^{out}-B]\rangle|-|\langle [M^{out}-A,B]\rangle|\}.
\end{eqnarray}
Here we assumed $[M^{out},B^{out}]=0$.
From this relation (25) one can derive, using the suitable variations of  Robertson's relation (20) such as $\sigma(M^{out}-A)\sigma(B)\geq \frac{1}{2}|\langle [M^{out}-A,B]\rangle|$,
\begin{eqnarray}
\{\sigma(M^{out}-A)+\sigma(A)\}\{\sigma(B^{out}-B)+\sigma(B)\}
\geq |\langle [A,B]\rangle|,
\end{eqnarray}
or 
\begin{eqnarray}
&&\sigma(M^{out}-A)\sigma(B^{out}-B)+\sigma(M^{out}-A)\sigma(B)+
\sigma(A)\sigma(B^{out}-B)\nonumber\\
&&\geq \frac{1}{2}\{|\langle [A,B]\rangle|,
\end{eqnarray}
while one obtains directly from (25)
\begin{eqnarray}
\sigma(M^{out}-A)\sigma(B^{out}-B)
\geq \frac{1}{2}|\langle [A,B]\rangle|,
\end{eqnarray}
if one assumes the unbiased measurement and disturbance using (23). 
The relations (26), (27) and (28) lead to the relations (1), (16) and (13),
respectively. This formulation shows that  the saturation of Robertson's relation (20) is a {\em necessary condition} of the saturation of our universally valid Heisenberg relation (1) and Ozawa's error-disturbance relation (16); in other words, the validity of Robertson's relation (20) ensures the validity of the relations (1) and (16). If one attaches importance to accuracy, the direct evaluation of the second-line of (25) is more accurate than the relations (1) or (16); the substantial deviation from the lower bound in Fig.2 suggests that the triangle inequality in (25) is not accurate.

One can also rewrite Robertson's relation (20) in the 
form
\begin{eqnarray}
&&\sigma(M^{out}-A)\sigma(B^{out}-B)\nonumber\\
&&\geq \frac{1}{2}|\langle-[M^{out},B^{out}]+\langle [M^{out}, B^{out}-B]+[M^{out}-A,B^{out}]\nonumber\\
&&\hspace{0.8cm} +[-A,-B]\rangle|\nonumber\\
&&\geq \frac{1}{2}\{|\langle [A,B]\rangle|-|\langle [M^{out},B^{out}-B]\rangle|-|\langle [M^{out}-A,B^{out}]\rangle|\nonumber\\
&&\hspace{0.8cm} -|\langle [M^{out},B^{out}]\rangle|\},
\end{eqnarray}
from which one obtains a universally valid relation (without assuming $[M^{out},B^{out}]=0$ in general),
\begin{eqnarray}
\{\sigma(M^{out}-A)+\sigma(M^{out})\}\{\sigma(B^{out}-B)+\sigma(B^{out})\}
\geq  \frac{1}{2}|\langle [A,B]\rangle|,
\end{eqnarray}
namely, another Heisenberg-type relation which may be compared to (26). In this relation  we do not assume the unbiased measurement and disturbance, and $\sigma(M^{out})$ and $\sigma(B^{out})$  stand for the ordinary standard deviations
which do not generally satisfy the relations (10) and (11). The relation (30) is amusing since it holds for any hermitian $M^{out}$ and $B^{out}$.
 
 All the  uncertainty relations discussed in this paper are thus derived from Robertson's relation (20) which is widely accepted as the mathematical expression of the Heisenberg uncertainty relation. From this point of view, all the universally valid relations are the secondary consequences of Robertson's relation (20), and one may pick up one of them on the basis of  its simplicity and usefulness. We prefer our universally valid Heisenberg relation (1) which provides support for the original idea of Heisenberg and exhibits an attractive similarity to the Arthurs-Kelly relation.
\\

The unbiased measurement which is defined by 
\begin{eqnarray}
\langle\psi\otimes\xi|M^{out}-A|\psi\otimes\xi\rangle=0 
\end{eqnarray}
for {\em all} $\psi$~\cite{appleby} is a specification of the measuring apparatus or procedure. This condition means that the deviations of the measuring apparatus from the true value of $A$ are balanced and vanish on average. In this sense it is a rather weak condition on the measuring apparatus. In comparison, the notion of unbiased disturbance is technically more subtle and one may suspect first the unbiased disturbance as the cause of the failure of the relations (9) and (13).

Another commonly used condition on the measurement, namely,
the precise measurement 
\begin{eqnarray}
\epsilon(A)=\langle\psi\otimes\xi|(M^{out}-A)^{2}|\psi\otimes\xi\rangle=0
\end{eqnarray}
imposes a condition on the measuring apparatus in (6), namely 
$(M^{out}-A)|\psi\otimes\xi\rangle=0$, for the hermitian operator $M^{out}-A$ and the positive-normed  Hilbert space. If the precise measurement is assumed to be valid for {\em all} $\psi$, then it would automatically imply the 
unbiased measurement (31). This is not logically natural since we test the notion of unbiased measurement in (13) or (21)  by using the precise measurement. We thus understand the precise measurement as follows: It implies that we can find a measuring 
procedure or apparatus which satisfies the relation (32) for any given state $\psi$ but 
the choice of the measuring procedure or apparatus  may generally depend on the given state
$\psi$. With this understanding, the precise measurement does not necessarily imply the unbiased measurement but we can still test the notion of unbiased measurement in  (13) or (21)  by using the notion of the precise measurement.     

A salient feature of Heisenberg's error-disturbance relation proposed by Ozawa (13) is that it does not contain explicitly the standard deviations which represent the intrinsic quantum fluctuations of the initial state.
This absence of standard deviations leads to a counter intuitive result that the repeated application of the {\em precise measurement} with $\epsilon(A)=0$ to a broadly spread initial state gives a broadly spread distribution of the measured values. The relation (13) contains only $\epsilon(A)$ which does not fully describe  the distribution of what the detector actually measures~\footnote{As a concrete example, the error $\epsilon(A)=0$ if one chooses the detuning angle $\phi=0$ in the spin measurement at Vienna~\cite{hasegawa}. 
The actual output of the detector of the {\em precise} measurement of $A=\sigma_{x}$ for the state $|\psi\rangle=|+z\rangle$ is $+1$ or $-1$ randomly with $\langle A\rangle=0$ and $\sigma(A)=1$. The experimental "error" used in the Heisenberg's error-disturbance relation of Ozawa (13) is $\epsilon(A)=0$ while the "error" used in the universally valid Heisenberg relation (1) is $\bar{\epsilon}(A)=\sigma(A)=1$. Our argument is that the use of the "measurement error" $\bar{\epsilon}(A)=\sigma(A)=1$ in the analysis of the uncertainty relation is more natural in the original spirit of Heisenberg.}. A similar comment applies to the "disturbance free" condition $\eta(B)=0$.
 The expression
\begin{eqnarray}
\sigma(M^{out})=\{\langle (M^{out}-A)^{2}\rangle+\langle (A-\langle A\rangle)^{2}\rangle \}^{1/2}
\end{eqnarray}
in the Arthurs-Kelly relation, for example, implies that the error $\epsilon(A)$ and the standard deviation $\sigma(A)$ are two independent physical quantities. It is well-known that the philosophy of Heisenberg is to express physical laws in terms of observed quantities,  and our conjecture is that the explicit presence of the standard deviation in some way, such as $\sigma(M^{out})$ in (33) or inaccuracy $\bar{\epsilon}(A)$ in (2), is essential for the formulation of satisfactory uncertainty relations\footnote{This issue is  related to the fundamental question what is the proper mathematical expression of "error" or "disturbance" which Heisenberg had in his mind in the consideration of his uncertainty relation.}. 
This observation may have some implications on the controversy of quantum limits on the measurement of gravitational waves~\cite{maddox}.

\section{Discussion and conclusion} 

We have presented a numerical illustration of the universally valid Heisenberg uncertainty relation (1) in comparison with  the modified Arthurs-Kelly relation (9) and the experimental data~\cite{hasegawa}. This illustration shows that the modified Arthurs-Kelly relation fails in the context of spin measurements but the universally valid Heisenberg relation is always valid and satisfactory in practical applications. It is hoped that the universally valid Heisenberg uncertainty relation, as a combination of the original idea of Heisenberg and a new ingredient added by  Ozawa, will play an important role in the future practical applications in view of the failure of the modified Arthurs-Kelly relation and the Heisenberg's  error-disturbance relation of Ozawa. In the semi-classical limit 
\begin{eqnarray}
\epsilon(A)\gg \sigma(A), \ \ \ \ \ \eta(B)\gg \sigma(B),\nonumber
\end{eqnarray}
 the modified Arthurs-Kelly
relation (9) and the Heisenberg's error-disturbance relation of Ozawa (13)
become consistent with our universally valid Heisenberg relation (1).

It was also shown that the essence of all the universally valid uncertainty relations is exhausted by Robertson's relation and, in particular, our universally valid Heisenberg uncertainty relation (1) is included in  Robertson's relation (20) which is widely accepted as the Heisenberg uncertainty relation. The use of the name "Heisenberg" in our relation is thus in line with its historical use.  The origin of the failure of the Heisenberg's  error-disturbance relations of Ozawa (13) was identified with  the assumptions of unbiased measurement and disturbance, and the invalidity of the assumption of unbiased disturbance was suspected as the main cause of the failure of (9) and (13). 

As for the experimental implications of the present study, it will be interesting to examine the Arthurs-Kelly relation itself for bounded observables such as spin variables
since the relation is then based on the unbiased measurements only and thus one can test the idea of the unbiased measurement independently of the unbiased disturbance. If the Arthurs-Kelly 
relation itself~\cite{arthurs,arthurs2}, 
\begin{eqnarray}
\sigma(M^{out})\sigma(N^{out})
&=&\{\langle (M^{out}-A)^{2}\rangle+\langle (A-\langle A\rangle)^{2}\rangle \}^{1/2}\nonumber\\
&&\times\{\langle (N^{out}-B)^{2}\rangle+\langle (B-\langle B\rangle)^{2}\rangle \}^{1/2}\nonumber\\
&\geq& |\langle [A,B]\rangle|
\end{eqnarray}
should fail, where $N^{out}$ stands for the meter variable for the variable $B$ with $[M^{out}, N^{out}]=0$, we would need to consider a universally valid version of the Arthurs-Kelly relation suggested in~\cite{fujikawa}
\begin{eqnarray}
\bar{\epsilon}(A)\bar{\epsilon}(B)&=&\{\langle (M^{out}-A)^{2}\rangle^{1/2}+\langle (A-\langle A\rangle)^{2}\rangle^{1/2} \}\nonumber\\
&&\times\{\langle (N^{out}-B)^{2}\rangle^{1/2}+\langle (B-\langle B\rangle)^{2}\rangle^{1/2} \}\nonumber\\
&\geq& |\langle [A,B]\rangle|.
\end{eqnarray}
See footnote [11] in~\cite{fujikawa}.

Finally, we mention an interesting analogy between the failure of the Heisenberg's  error-disturbance relations of Ozawa (13) and the subtle behavior of the Kennard-Robertson relation for the periodic boundary condition in box normalization. It is known that complications are associated with the uncertainty relations for periodic systems~\cite{judge, judge-lewis, tanimura}. The one-dimensional Schr\"{o}dinger problem with the periodic boundary condition $\psi(-L/2,t)=\psi(L/2,t)$ in a box $[-L/2, L/2]$ gives the universally valid Kennard-Robertson relation~\cite{fujikawa2}\footnote{The one-dimensional Schr\"{o}dinger problem implies
\begin{eqnarray}
\int_{-L/2}^{L/2} dx|[is(p-\langle p\rangle)+(x-\langle x\rangle)]\psi(x,t)|^{2}\geq 0,\nonumber
\end{eqnarray}
where we consider the wave function which satisfies the periodic condition $\psi(-L/2,t)=\psi(L/2,t)$.
Here $s$ is an arbitrary real parameter. This relation is re-written after partial integration 
\begin{eqnarray}
&&\int_{-L/2}^{L/2} dx\psi(x,t)^{\star}[-is(p-\langle p\rangle)+(x-\langle x\rangle)]\nonumber\\
&&\hspace{ 1 cm}\times[is(p-\langle p\rangle)+(x-\langle x\rangle)]\psi(x,t)+s\hbar L|\psi(L/2,t)|^{2}\nonumber\\
&&=s^{2}\langle (p-\langle p\rangle)^{2}\rangle + 
\langle (x-\langle x\rangle)^{2}\rangle +s \{\hbar L|\psi(L/2,t)|^{2}-i\langle [p,x]\rangle\} \geq 0,\nonumber
\end{eqnarray}
where the extra term $s\hbar L|\psi(L/2,t)|^{2}$ arises since 
$x\psi(x,t)$ may not be periodic even for periodic $\psi(x,t)$ with $\psi(-L/2,t)=\psi(L/2,t)$. This quadratic relation in $s$ holds for any real $s$, and thus the discriminant gives
eq.(36). An alternative derivation of (36) is found in~\cite{fujikawa2}.  See also~\cite{tanimura}.} 
\begin{eqnarray}
\langle (p-\langle p\rangle)^{2}\rangle^{1/2} 
\langle (x-\langle x\rangle)^{2}\rangle^{1/2} \geq \frac{1}{2}\hbar|1-L|\psi(L/2,t)|^{2}|,
\end{eqnarray}
which does not appear to be widely recognized. The pure plane wave solution with the periodic boundary condition in a finite box, which gives $\Delta(p)=\langle (p-\langle p\rangle)^{2}\rangle^{1/2}=0$ for the state with a discrete momentum eigenvalue and a finite value for $\Delta(x)=\langle (x-\langle x\rangle)^{2}\rangle^{1/2}$, is consistent with (36) since the right-hand side of (36) also vanishes for the pure plane wave. This is analogous to (21).
On the other hand, if one assumes that the right-hand side of the Kennard-Robertson relation (36) is independent of the state
vector $\psi$ (in analogy with the ordinary definition of {\em unbiased} measurement~\cite{appleby}), the pure plane wave solution with the periodic boundary condition in a finite box then contradicts the uncertainty relation (36) with the lower bound replaced by the conventional $\hbar/2$. This is analogous to the failure of the Heisenberg's error-disturbance relation of Ozawa (13) for bounded operators such as spin variables. This  complication does not arise for the Kennard-Robertson relation defined for the normalizable states in an open space $[-\infty, \infty]$, which implies $L|\psi(L/2,t)|^{2}\rightarrow 0$ for $L\rightarrow\infty$.

\end{document}